# Antiproton production with a fixed target and search for superheavy particles at the LHC

A. B. Kurepin[1*], N. A. Kurepin[2], K. A. Skazytkin[2]

[1] Institute for Nuclear Research, 117312 Moscow, Russia

[2] Skobeltsyn Institute of Nuclear Physics, Lomonosov Moscow State University, 119991, Moscow, Russia

**Abstract**

A proposal for an experiment to measure the cross section of antiproton production in a proton-nuclear collision in a kinematically forbidden region for nucleon-nucleon interaction on a fixed LHC target is considered. It is shown that this process can be separated from the kinematically allowed production process using the existing detectors of the ALICE facility at a proton energy of 7 TeV with a fixed nuclear target. Assuming the scale dependence of the cross section, the data obtained can be used to estimate the subthreshold cross section for the production of superheavy particles with a mass of several tens of TeV in the LHC lead nucleus beam.

## 1. Introduction

Among the processes of particle production in nuclear collisions at high energy beams, one can highlight those where production occurs under kinematic conditions that are forbidden in the nucleon-nucleon interaction. At present, this problem is of great importance for studying the possibility of particle production with beams of lead nuclei for a mass exceeding 14 TeV, which is attainable in a nucleon - nucleon collision at the LHC. Actually, the energy at the center of mass in the collision of lead nuclei at the LHC at a beam energy of 2.76 TeV per nucleon is 1150 TeV. The existence of such superheavy particles with a mass much less than the Planck mass is forbidden in the Grand Unification Theory. However, when trying to solve the problem of the hierarchy of interactions in models of superstrings, super symmetry and introducing additional dimensions, the creation of such particles is allowed. For example, in the model of extra dimensions with a compactification radius of one Fermi, particles with a mass of several tens of TeV can be produced [1-2].

Obviously, for the production of superheavy particles, the cross section of the "subthreshold" process will be very small. Theoretical estimates of this cross section are apparently impossible at the present time. However, an analogy can be drawn with the subthreshold production of antiprotons at intermediate energies, since the subthreshold process is obviously associated with the correlation of nucleons or quarks in colliding nuclei and could be weakly dependent on energy. Nevertheless, such a study must also be carried out at higher energies.

---

*e-mail: kurepin@inr.ru

This investigation can be carried out in the study of the production process under kinematic conditions that are forbidden in the nucleon-nucleon interaction. Such a study was started at the U-70 accelerator in Protvino [3] at an energy of 19 GeV ($\sqrt{s}$ = 6.05 GeV) per nucleon on carbon nuclei. As shown later in this article, it is possible to measure the production of antiprotons in a proton beam with an energy of 7 TeV at the LHC on a fixed target of heavy nuclei ($\sqrt{s}$ = 114.7 GeV).

## 2. Subthreshold production of antiprotons in nuclear collisions

The study of the production of antiprotons in proton-nuclear and nucleus-nucleus collisions at energies below the production threshold in a nucleon-nucleon collision has been the subject of a significant number of works [4 - 8]. The measurements were carried out with proton and nuclear beams at JINR, BNL, GSI and KEK. In all experiments, the values of production cross sections were obtained, which significantly exceed the estimates obtained when taking into account the lowering of the production threshold due to the Fermi motion of nucleons in the nucleus [9]. A unified phenomenological description of all experimental data was obtained in the generalized parton model [10-11]. The model is based on the introduction of the parton distribution parameter not only in the target nucleus (x), but also in the incident nucleus (z). Due to the conservation of the 4-momentum, we obtain:

$$( zP_1 + x P_2 - P_d )^2 = ( zP'_1 + x P'_2 + P'_i )^2 \qquad (1)$$

where $P_1$, $P_2$, $P_d$ - are the 4-momenta of the incident and fixed nucleons in the nuclei and of the produced antiproton, $P_i$ is the 4-momentum of an additional proton for a conservation of the baryon number. For the maximum values of the momentum of the produced antiproton:

$$\vec{p}'_1 = \vec{p}'_2 = \vec{p}'_i = 0 \qquad (2)$$

we have: $(P'_1)^2 = m^2$, $(P'_2)^2 = m^2$, where m – is the mass of a nucleon, $\vec{p}'_1$, $\vec{p}'_2$, $\vec{p}'_i$ - momenta of an incident and fixed nucleons in nuclei and an additional proton.

From equation (1), the mutual dependence of the parameters x and z can be obtained [10]. The production of antiprotons in nucleus-nucleus collisions now is possible not only at small values of the parton parameters, but also in the kinematically forbidden region for nucleon - nucleon interactions at x > 1 and z > 1.

In the article [11], a universal dependence of the scaling type of all subthreshold data on the production of antiprotons was obtained at x in the range 1 – 4 and at z values equal to 1 for incident protons, 1.3 for deuterons, 2 for carbon nuclei, 3 for more heavy nuclei at energies from 2 to 6 GeV per nucleon:

$$( A_1 A_2 )^{-0.43} \cdot E_1 \frac{d^3 \sigma}{dp^3} \ [ \text{mb GeV}^{-2} \ c^3 \ \text{sr}^{-1} ] = 0.57 \exp( -x/0.158 ) \qquad (3)$$

This dependence has an exponential form and describes well the data for proton - nuclear collisions. For nucleus - nucleus collisions, the deviations of individual data from the curve are larger, but the range of cross sections reaches four orders of magnitude.

Assuming a weak dependence of the obtained dependence of the subthreshold production cross section on energy, the yield of superheavy particles with a mass of 16 TeV on the LHC proton beam in the collision of lead nuclei was estimated [12]. The obtained value of about 70 particles per year allows planning the corresponding experiment. However, for more reasonable estimates, it is necessary to determine the dependence of the production cross section on the scaling parameters in the subthreshold process, i.e. at values $x > 1$, at high energies more close to the LHC energies.

In the next section, we analyze the possibility of measuring the cross section for antiproton production in the kinematically forbidden region on a fixed target of the LHC collider.

### 3. Production of antiprotons in a kinematically forbidden region at a fixed target of the LHC collider.

To determine the possibility of measuring the yield of antiprotons in the kinematically forbidden region on a fixed target of the LHC collider [13] the calculation was performed of the kinematics of antiproton production at an energy of 7 TeV on bismuth nuclei. In Fig. 1 the magnitudes of the maximum transverse momentum values of antiprotons in the center-of-mass system are given for the parameter $x = 1$ and $x = 2$, depending on the pseudo-rapidity. It can be seen that these magnitudes differ significantly. Thus, it is possible to separate the kinematically allowed process from the forbidden in the nucleon-nucleon collision.

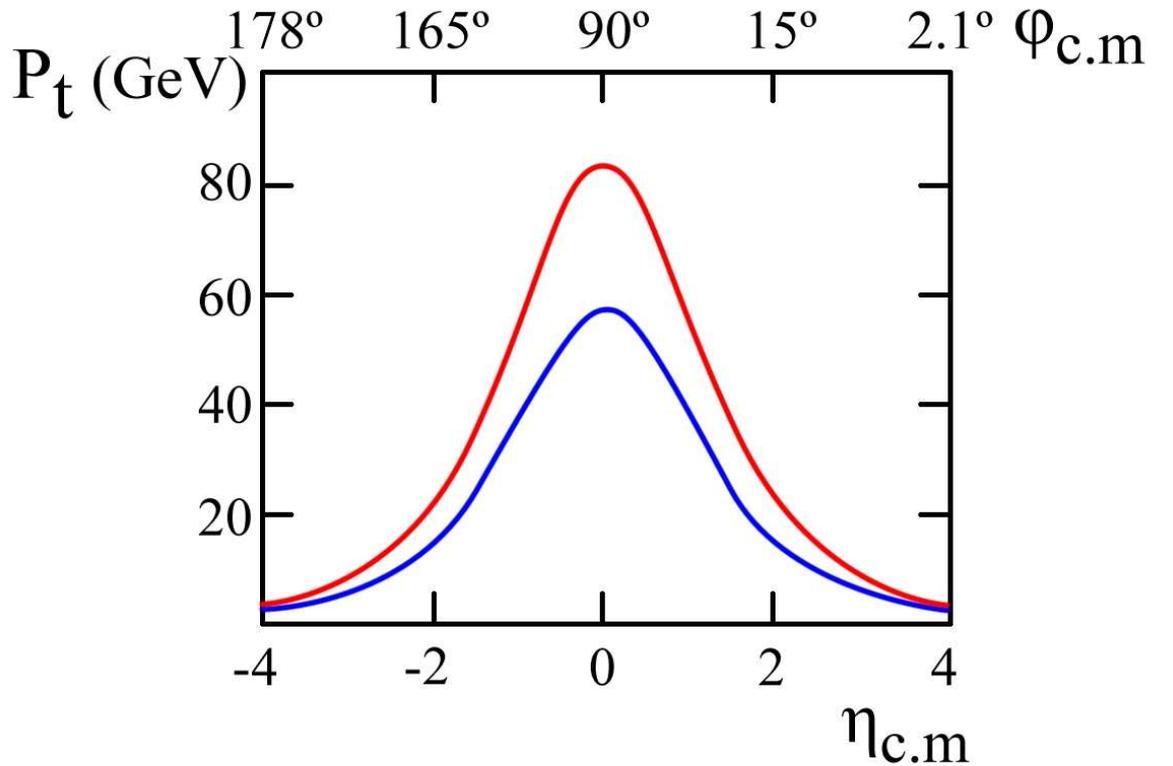

Fig. 1. Dependence of the transverse momentum of antiprotons on the rapidity in the center of mass. Blue line with $x = 1$, red line with $x = 2$.

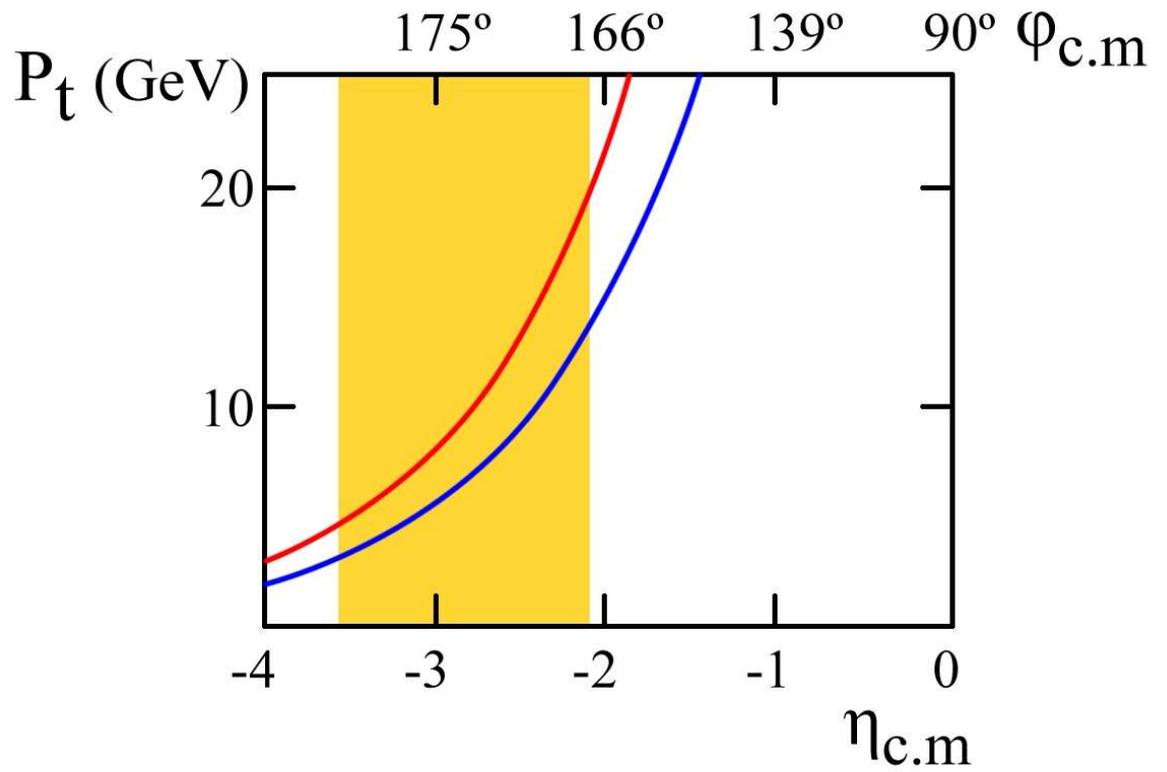

Fig. 2. Dependence of the transverse momentum of antiprotons on the rapidity in the center of mass. Blue line with x = 1, red line with x = 2. The area available with a fixed target and x = 1 is marked in yellow.

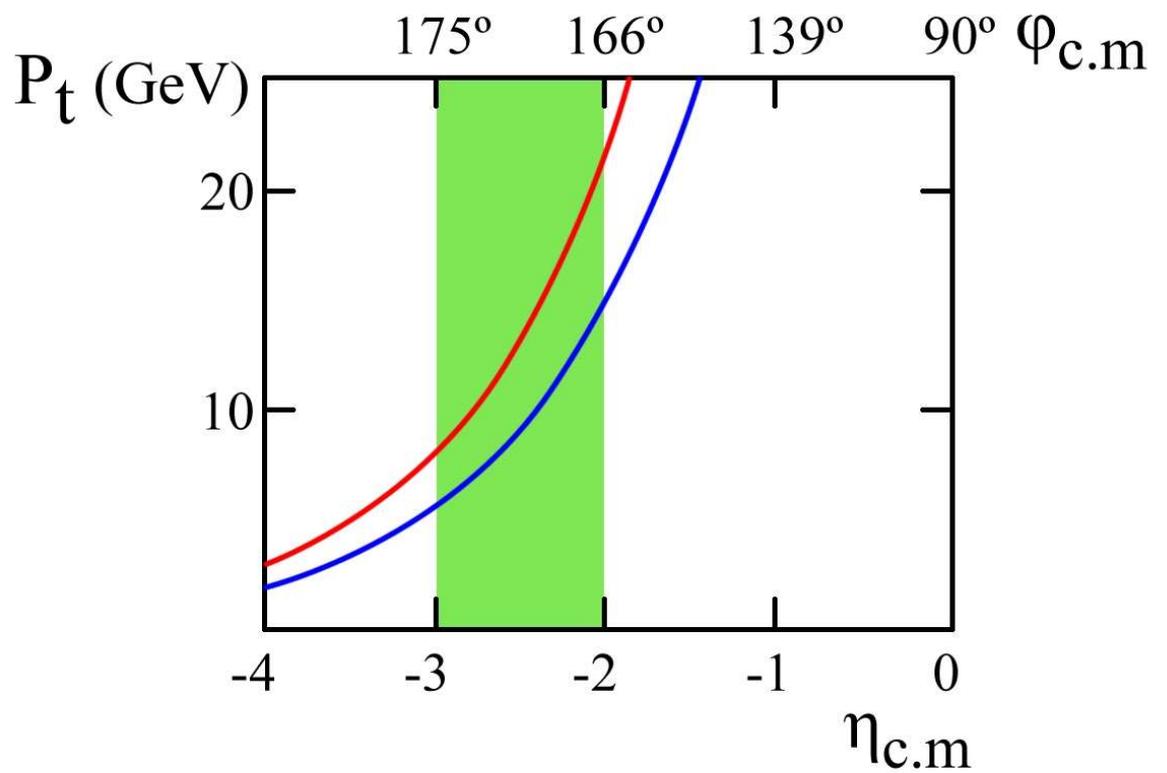

Fig. 3, Dependence of the transverse momentum of antiprotons on the rapidity in the center of mass. Blue line with x = 1, red line with x = 2. The area available with a fixed target and x = 2 is marked in green

The identification and measurement of the transverse momentum in the ALICE installation is carried out by the TPC projection camera. The fixed target will be located at a distance of 480 cm from the IP. The geometry of the TPC and the beam tube limits the range of possible angles from 5 to 28 degrees. The Fig. 2 and 3 show the corresponding intervals of pseudo-rapidity for x = 1 and x = 2. The required transverse momentum ranges of 3 - 20 GeV are available for TPC measurements [14]

In the laboratory system, there is an ambiguity in the dependence of the transverse momentum on the pseudo-rapidity (Fig. 4). However, antiprotons with a small transverse momentum can be easily separated.

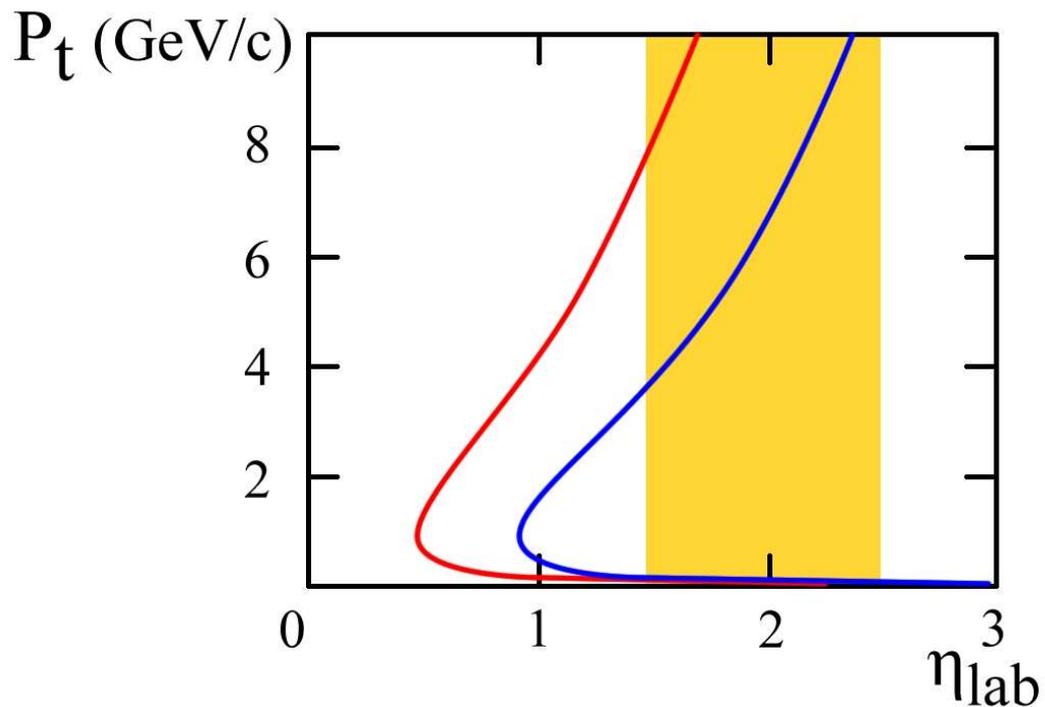

Fig. 4. Dependence of the transverse momentum of antiprotons on the rapidity
in a laboratory system. Blue line with x = 1, red line with x = 2.
The area available with a fixed target is marked in yellow.

When analyzing the measured antiproton production cross sections, parameters x and z must satisfy the following relationship:

$$x = \frac{z \cdot E \cdot E_d \cdot (1 - \cos(\theta)) - M^2}{(z \cdot E - E_d)m} \qquad (4)$$

where $E$ is the beam energy, $E_d$ - is the antiproton energy, $M$ is the antiproton mass, $m$ is the nucleon mass, $\theta$ is the production angle.

At a high energy of 7 TeV of a proton beam, the parameter x is practically independent on the parameter z and on the beam energy:

$$x \approx \frac{E_d \cdot (1 - \cos\ )}{m} \qquad (5)$$

For the parameters of the planned experiment: $\theta = 28^0$, $\Delta p = 1$ GeV, $\Delta \Omega = 0.1$ sr the values of parameters x, as a function of the antiproton transverse momentum, are given in Table 1. The same table shows the production cross section calculated by formula (3) and the antiproton yield at a luminosity of $10^{30}$ см$^{-2}$ сек$^{-1}$

Table 1

Parameter x, antiproton production cross sections and antiproton yield as a function of the antiproton transverse momentum

| $P_t$ | 4 | 6 | 8 | GeV |
|---|---|---|---|---|
| $E_d$ | 8.5 | 12.8 | 17 | GeV |
| X | 1.1 | 1.6 | 2.18 | - |
| $\sigma_{inv}$ | $8 \cdot 10^{-3}$ | $6 \cdot 10^{-4}$ | $8 \cdot 10^{-6}$ | $mb\ GeV^{-2} c^3 sr^{-1}$ |
| $N_d$ | $25 \cdot 10^3$ | $3 \cdot 10^3$ | 50 | 1/hour |

## 4. Conclusions

Investigation of antiproton production in the kinematically forbidden region in the nucleon-nucleon interaction on a fixed target of the LHC collider is possible with the existing detectors of the ALICE facility. The data obtained on the dependence of the subthreshold production cross section on the scaling parameter x > 1 can be used to estimate the yield of superheavy particles production with the LHC lead nucleus beam.

## 5. Acknowledgement

We thank the members of the ALICE fixed target study group and of the AFTER study group for useful discussions and the interest in the work.